# Chemical Bonding of Termination Species in 2D Carbides Investigated through Valence Band UPS/XPS of $Ti_3C_2T_x$ MXene

Lars-Åke Näslund[1], Mikko-Heikki Mikkelä[2], Esko Kokkonen[2] and Martin Magnuson[1]

[1] Thin Film Physics Division, Department of Physics, Chemistry, and Biology (IFM), Linköping University, SE-581 83 Linköping, Sweden.
[2] MAX IV Laboratory, Lund University, SE-221 00 Lund, Sweden.

E-mail: lars-ake.naslund@liu.se

**Abstract**

MXenes are technologically interesting 2D materials that show potential in numerous applications. The properties of the MXenes depend at large extent on the selection of elements that build the 2D MX-layer. Another key parameter for tuning the attractive material properties is the species that terminate the surfaces of the MX-layers. Although being an important parameter, experimental studies on the bonding between the MX-layers and the termination species are few and thus an interesting subject of investigation. Here we show that the termination species fluorine (F) bonds to the $Ti_3C_2$-surface mainly through Ti $3p$ – F $2p$ hybridization and that oxygen (O) bonds through Ti $3p$ – O $2p$ hybridization with a significant contribution of Ti $3d$ and Ti $4p$. The study further shows that the $Ti_3C_2$-surface is not only terminated by F and O on the threefold hollow face-centered-cubic (fcc) site. A significant amount of O sits on a bridge site bonded to two Ti surface atoms on the $Ti_3C_2$-surface. In addition, the results provide no support for hydroxide (OH) termination on the $Ti_3C_2$-surface. On the contrary, the comparison of the valence band intensity distribution obtained through ultraviolet- and x-ray photoelectron spectroscopy with computed spectra by density of states, weighed by matrix elements and sensitivity factors, reveals that OH cannot be considered as an inherent termination species in $Ti_3C_2T_x$. The results from this study have implications for correct modeling of the structure of MXenes and the corresponding materials properties. Especially in applications where surface composition and charge are important, such as supercapacitors, Li-ion batteries, electrocatalysis, and fuel- and solar cells, where intercalation processes are essential.

## 1. Introduction

MXenes ($M_{n+1}X_nT_x$, n = 1, 2, or 3) are a family of two-dimensional (2D) transition metal carbides and nitrides that show promising findings within a wide range of potential applications such as supercapacitors [1], Li-ion batteries [2], fuel- and solar cells [3], transparent conductive electrodes [4], and composite materials with high strength [5]. Important for the material properties that make a MXene suitable for a particular application are the atoms and small molecules that terminate the surfaces of the 2D $M_{n+1}X_n$- flakes. Through selected combination of a transition metal (M), a carbon or nitrogen (X), numerical proportions of atoms (n), and termination species ($T_x$) the $M_{n+1}X_nT_x$ can be tailor designed for a specific application with fine-tuned properties.

MXenes are synthesized by exfoliation of the parent precursor compounds MAX phases [6], which are nanolaminated phases of carbides and/or nitrides following the general formula $M_{n+1}AX_n$ (n = 1, 2, or 3) where A is a group A element in the Periodic Table [7]. To produce MXene the MAX phase is immersed in an acidic solution where the weakly bonded A-layers [8] are removed though selective etching [9,10]. The termination species that cover the surfaces of the newly formed 2D $M_{n+1}X_n$-layers depends on the selected etchant [11-13]. The most studied MXene so far is $Ti_3C_2T_x$ where the inherent $T_x$ suggests being oxygen (O) [8,14-16], fluorine (F) [8,14-16], and hydroxide (OH) [15,16].

The obtained $Ti_3C_2T_x$ samples in this study consist of a high number of flakes stacked one on top of the others to form a thin film. Each flake is a $Ti_3C_2$-layer that consists of three Ti-monolayers and two C-monolayers in an alternating sequence where the first and the last Ti-monolayers form surfaces that are terminated by, e.g., O and F that occupies the threefold hollow face-centered-cubic (fcc) sites [14], formed by three surface Ti atoms above a middle monolayer Ti atom as shown in figure 1. A recent study has found that in addition to the fcc-site the O can also occupy other sites that are not yet identified [14], although a bridge site between Ti and C atoms is proposed [17]. The chemical bonding of the termination species to the $Ti_3C_2$-surface is one of the most important parameters because $T_x$ modification is the key when to fine-tune the properties of MXene materials [13,18]. Different theoretical and experimental results, which are not always consistent, show the importance of the need of more understanding regarding the bonding conditions at the surfaces of the MXenes [13].

The aim of this study is to present and assign the fine structures in the valence band of well-defined and near impurity-free $Ti_3C_2T_x$ using ultraviolet- and x-ray photoelectron spectroscopy (UPS/XPS) and resonant photoemission spectroscopy (PES) at photon energies of 27-80, 780, and 1486.6 eV by utilizing a combination of synchrotron- and Al-Kα radiation. Features in a valence band UPS/XPS spectrum are very sensitive to many different parameters, such as molecular orbital interactions, photoionization cross-sections, resonant enhancements, and core screening processes [19]. It is therefore very seldom possible to assign obtained features directly to components, e.g., to the termination species O, F, and OH. However, by employing light with different photon energies it is possible to reduce or enhance structures that originate from separate hybridized orbital states because the UPS/XPS/PES intensity at different photon energies varies with the photoionization cross-sections and resonant phenomena [19,20]. Since the valence band structures are very sensitive to the local environment of the probed species, the photon energy dependent UPS/XPS/PES will provide valuable information regarding bonding conditions as well as structural arrangement of termination species in comparison to a recent Ti K-edge XAS study [21].

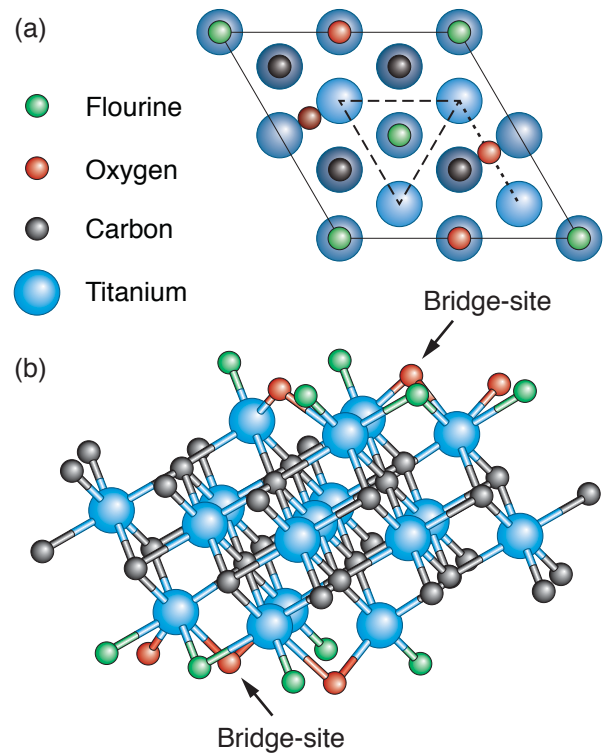

**Figure 1.** Crystal structure of $Ti_3C_2$ terminated by F and O occupying fcc-sites and O also occupying a bridge site between two surface Ti atoms, one O on each side of the $Ti_3C_2$-layer. (a) Top view where the dashed triangle highlights the fcc-site occupied by an F atom and the dotted line highlights the bridge site, between two surface Ti atoms, occupied by an O atom. Darker Ti and O atoms are in the middle and bottom monolayers. (b) Side view showing that F and O atoms terminate both sides of the $Ti_3C_2$-layer.

## 2. Methods

An earlier work has shown that oxidized sections of MXene samples are detrimental for $Ti_3C_2T_x$ valence band studies as the O 2*p* states of oxidized Ti dominate the central part of the valence band binding energy region, i.e., 5-10 eV below the Fermi level ($E_F$) [22]. It is therefore a necessity to produce $Ti_3C_2T_x$ samples that are free from oxide formation. In the present study a thin film $Ti_3C_2T_x$ sample was produced through HF(aq) etching of a well-defined thin film $Ti_3AlC_2$ obtained from direct current magnetron sputtering of Ti, Al, and C targets in an ultra-high vacuum (UHV) environment [14]. In addition, a freestanding film of $Ti_3C_2T_x$ produced through immersion of $Ti_3AlC_2$ powders in an aqueous solution of hydrochloric acid (HCl) and lithium fluoride (LiF) was also included [1]. In both cases the preparation of the $Ti_3AlC_2$, the Al removal through etching, and the storage of the obtained $Ti_3C_2T_x$ samples in vacuum or inert argon (Ar) atmosphere, occurred within a couple of hours and with minimum exposure to the ambient environment. Impurities, such as $TiO_2$ and $Al_2O_3$, and contamination, such as hydrocarbon and alcohol compounds, could therefore be kept at an insignificant level [8].

### 2.1 $Ti_3AlC_2$ thin film preparation

A $Ti_3C_2T_x$ thin film on substrate was prepared from $Ti_3AlC_2$ deposited on a c-axis-oriented sapphire ($Al_2O_3$) plate. The 10x10 $cm^2$ sapphire plate (MTI Corp., CA) was placed on a rotating substrate holder and preheated at 780 °C for 1 h in a UHV environment before deposition of Ti, C, and Al using direct current magnetron sputtering. The three elemental targets Ti, C, and Al with diameters of 75, 75, and 50 mm, respectively, were ignited with powers of 92, 142, and 26 W, respectively. With the substrate temperature kept at 780 °C an incubation layer of TiC was formed on the $Al_2O_3$ substrate by igniting the Ti and C targets 30 s before igniting the Al target. The sputtering process gas was Ar (99.9999 % purity) at a constant pressure of about 5 mbar. The co-sputtering of the Ti, C, and Al targets occurred for 10 min and produced a uniform $Ti_3AlC_2$ film about 30 nm thick. Before the thin film $Ti_3AlC_2$ sample was removed from the deposition system the temperature was cooled down to near room temperature (RT).

### 2.2 $Ti_3AlC_2$ powders preparation

A sample of freestanding $Ti_3C_2T_x$ film was prepared from $Ti_3AlC_2$ powders. The $Ti_3AlC_2$ powders were produced from a mixture consisting of TiC (Alfa Aesar, 98+ %), Ti (Alfa Aesar, 98+ %), and Al (Alfa Aesar, 98+ %) in a 1:1:2 molar ratio. The mixture was processed in a mortar with a pestle for 5 min and thereafter inserted into an alumina tube furnace. With a continuous stream of Ar gas the furnace was heated at a rate of 5 °C $min^{-1}$ up to 1450 °C and held for 280 min before cooling down to room temperature. The obtained $Ti_3AlC_2$, which was lightly sintered, was thereafter crushed to powders with particle sizes < 60 μm using a mortar and pestle.

### 2.3 $Ti_3C_2T_x$ thin film on substrate preparation

To minimize the exposure to the atmosphere the $Ti_3AlC_2$ thin film was converted into $Ti_3C_2T_x$ immediately after the $Ti_3AlC_2$ thin film deposition. The $Ti_3AlC_2$ thin film was etched in 10 % concentrated HF(aq) (Sigma Aldrich, Stockholm, Sweden) for 60 min at room temperature. After etching, the sample was rinsed in deionized water and ethanol and without delay inserted into an Al-Kα XPS system. Impurities, such as $TiO_2$ and $Al_2O_3$, and contamination, such as hydrocarbon and alcohol compounds, could therefore be kept at an insignificant level.

### 2.4 Freestanding $Ti_3C_2T_x$ film preparation

Also the $Ti_3AlC_2$ powders was converted into $Ti_3C_2T_x$ immediately after the $Ti_3AlC_2$ powders with particle sizes < 60 μm was formed. Half a gram of the obtained $Ti_3AlC_2$ powders was added to a premixed 10 mL aqueous solution of 12 M HCl (Fisher, technical grade) and 2.3 M LiF (Alfa Aesar, 98+ %) in a Teflon bottle placed in an ice bath to avoid the initial overheating. After 0.5 h in the ice bath the bottle was placed in a 35 °C oil bath for 24 h while continuous stirring. The mixture was thereafter washed through three cycles of 40 mL 1 M HCl(aq), three cycles of 40 mL 1 M LiCl (Alfa Aesar, 98+ %), and several cycles of 40 mL deionized water until the supernatant reached a pH of approximately 6. The mixture was diluted with 45 mL of deionized water, de-aerated by bubbling $N_2$ gas, and sonicated using an ultrasonic bath for 60 min. The resulting suspension was centrifuged for 60 min at 3500 rpm, which removed larger particles. The supernatant produced had a $Ti_3C_2T_x$ concentration of 1 g $L^{-1}$. To make freestanding films, 20 mL of the suspension was filtered through a nanopolypropylene membrane (3501 Coated PP, 0.064 μm pore size, Celgard, USA).

The freestanding $Ti_3C_2T_x$ film was immediately stored in Ar atmosphere and within 24 hours transported to the MAX IV synchrotron radiation facility and placed in an XPS endstation. Hence, impurities, such as $TiO_2$ and $Al_2O_3$, and contamination, such as hydrocarbon and alcohol compounds, could be kept at an insignificant level.

### 2.5 UPS/XPS/PES measurements

The UPS/XPS/PES measurements of the $Ti_3C_2T_x$ samples were performed using both an in-house Al-Kα radiation XPS system at Linköping University (Linköping, Sweden) and a synchrotron radiation UPS/XPS system at MAX IV (Lund, Sweden). The UPS/XPS spectra could therefore be obtained using photon energies of 27-1486.6 eV. The benefit of using different photon energies is the variation in the photoionization cross-sections at the selected photon energies [20], which will assist in the analysis of the valence band features. For confirmation of sample quality, e.g. the ratio of termination species and the amount of impurities and contaminations, the F 1$s$, O 1$s$, Ti 2$p$, C 1$s$, and Cl 2$p$ core-levels were recorded in addition to the valence band XPS measurements.

The in-house XPS was performed using an AXIS Ultra[DLD] surface science system from Kratos Analytical Ltd. (Manchester, U.K.). The x-ray beam (monochromatic Al-Kα radiation) irradiated the surface of the sample at an angle of 45°, with respect to the surface normal, and provided an x-ray spot size of 300 x 800 μm$^2$. The electron energy analyzer accepted the photoelectrons perpendicular to the sample surface with an acceptance angle of ±15°. A series of core-levels and valence band XPS spectra were recorded with a step size of 0.1 eV and pass energy of 20 eV, which provided an overall energy resolution better than 0.30 eV.

The synchrotron radiation-based (SR) UPS/XPS were obtained using a SPECS Phoibos 150 NAP electron energy analyzer installed on the ambient pressure XPS endstation at the SPECIES beamline [23,24]. The sample was irradiated at an angle of 57°, with respect to the surface normal, and provided an x-ray spot size of 100 x 100 μm$^2$. The electron energy analyzer accepted the photoelectrons perpendicular to the sample surface with an acceptance angle of ±22°. The core-levels and valence band UPS/XPS spectra were recorded with a step size of 0.05 eV and pass energy of 50 eV, which provided an overall energy resolution better than 0.25 eV.

In both XPS systems used in this study, the sample heating was performed in a separated preparation chamber. In the in-house XPS system the Ti$_3$C$_2$T$_x$ thin film on the sapphire substrate was initially heated to 200 °C to remove the adventitious carbon and after the first set of XPS measurements a series of heating and XPS measurements was performed where the temperature range was 500-750 °C with 50 °C steps and 30 min dwell time at each heating cycle. In the SR-XPS endstation at the SPECIES beamline the first set of UPS/XPS measurements on the freestanding Ti$_3$AlC$_2$ film were performed before any heat treatments and a second set of UPS/XPS measurements were performed after heat treatment at 670 °C for 30 min. The sample temperature was about 100 °C when the Ti$_3$C$_2$T$_x$ samples were transferred in vacuo from the preparation chamber to the analyzer chamber in both XPS systems. The base pressures in both XPS systems were below 1x10$^{-9}$ mBar.

The binding energy scale of all core-level XPS and valence band UPS/XPS spectra was calibrated using E$_f$, which was set to a binding energy of 0.00 ±0.02 eV. The core-level spectra were normalized at the background on the low binding energy side of the main peak/peaks. The valence band spectra had the intensity of the higher order reflections from the monochromator subtracted, i.e. the spectra were subtracted by a constant obtained from the intensity at about 2 eV above E$_f$. The valence band spectra were thereafter normalized at the background at 16 eV below E$_f$.

*2.6 Valence band spectrum computations*

First-principle calculations were carried out using density functional theory (DFT) implemented in the Vienna ab initio simulation package (VASP) [25]. For the self-consistent calculations the projected augmented wave (PAW) method [26] was used with the Perdew-Burke-Ernzerhof (PBE) and the Heyd-Scuseria-Ernzerhof (HSE06) functionals of the generalized gradient approximation (GGA) [27]. The HSE06 functional is a hybrid functional including a linear combination of short- and long-range PBE exchange terms and a short-range Hartree-Fock term that improves the formation energies and band gaps [28]. The screening parameter of HSE was 0.2 Å$^{-1}$ with a plane wave cutoff energy of 700 eV. As the computational cost is significantly higher for the HSE06, in comparison to the PBE functional, a 4x4x1 k-grid was used for the HSE06 functional for the structure relaxation while a 31x31x3 k-grid was used for the PBE functional.

The structural models were 2D slabs that contained 17 Ti atoms, 8 C atoms, and 0-16 T$_x$ species (F, O, and OH). The distances between Ti and C in the 2D slab models were based on the bond length determination in a previous Ti K-edge extended x-ray absorption fine structure (EXAFS) study [21]. The distances between the termination species and the surface Ti atoms in the Ti$_3$C$_2$-layer were optimized to provide acceptable agreement with the experimental UPS/XPS valence band spectra. The density of states (DOS) distribution in the valence band region was weighted by the transition matrix elements and tabulated sensitivity factors using the Galore software [29]. The final state lifetime broadening was accounted for by a convolution with an energy-dependent Lorentzian function with a broadening increasing linearly with the distance from E$_f$ according to the function a+b(E−E$_f$), where the constants a and b were set to 0.025 eV and 0.055 (dimensionless), respectively.

## 3. Results and discussion

*3.1 Core-level XPS*

Series of the F 1*s*, O 1*s*, Ti 2*p*, and C 1*s* XPS spectra of a thin film Ti$_3$C$_2$T$_x$ on a sapphire substrate is presented in figure 2. The spectra evolve with the temperature and it is clear that the F 1*s* intensity decreases with the temperature and that O 1*s* shows an intensity redistribution that is synchronized with the F desorption. Further, the Ti 2*p* shows an intensity redistribution that also is synchronized with the F desorption while C 1*s* remains non-affected. The analysis of the temperature dependent F 1*s*, O 1*s*, Ti 2*p*, and C 1*s* XPS spectra, including an interpretation, is presented in details elsewhere together with a scanning tunneling electron microscopy (STEM) study [14]. In short, Persson et al. concluded that F and O both occupy the fcc-site, that F has precedence, and that O transfers from an alternative site to the fcc-site when the latter is available after F desorption. That C 1*s* XPS shows no changes upon F desorption or site redistribution of O indicates that the C atoms are not directly involved in the bonding of the terminating species. Thus, the consistent C 1*s* orbital rules out the proposed bridge site between Ti and C atoms or a hexagonal close packing (hcp) site, i.e., the hollow site of three Ti atoms above a C atom, as the alternative site for O. In addition, Persson et al. did not found any clear indications of OH being a termination species.

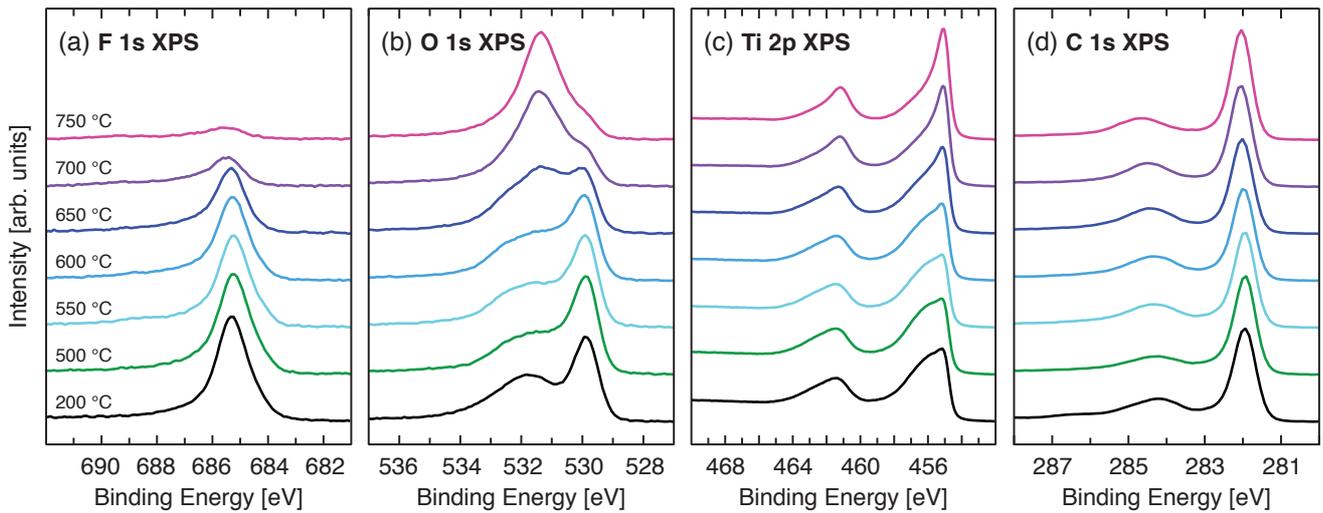

**Figure 2.** XPS spectra of $Ti_3C_2T_x$ for the core levels (a) F 1*s*, (b) O 1*s*, (c) Ti 2*p*, and (d) C 1*s*. The spectra of the $Ti_3C_2T_x$ sample heated to 200 and 500 °C show differences only for the O 1*s* XPS indicating a redistribution of O between different sites. Between 500 °C and 750 °C, in steps of 50 °C, there are significant changes for all spectra except for C 1*s*. The photon energy was 1486.6 eV.

### 3.2 Valence band XPS

Figure 3 shows the temperature-programmed XPS of the valence band region of the thin film $Ti_3C_2T_x$ sample as obtained using the photon energy 1486.6 eV (Al-K$\alpha$ radiation). The features in the binding energy region closest to $E_f$ (0-4 eV) do not show any significant changes during the heat treatment and can therefore be associated with the C 2*p* hybridization with Ti 3*p*-states, similar as in cubic titanium carbide (TiC) [30]. In addition, the broad feature around 11.3 eV, which is associated with the C 2*s* hybridization with the Ti 3*p*-states [30], remains unaffected during the heat treatment. From the temperature-programmed XPS of both the C 1*s* core level and the valence band it is clear that the C atoms in the $Ti_3C_2$-layer are not affected by the heat treatment. Hence, the C orbitals are only involved in the bonding towards the neighboring Ti atoms in the $Ti_3C_2$-layer, which remains stable up to at least 750 °C. The intensity redistribution at the center of the valence band shows, on the other hand, an activity at the surfaces of the $Ti_3C_2$-layers. As the temperature is raised above 550 °C the feature at 8.8 eV decreases and has vanished completely at 750 °C. Also the intensity at 5.4 eV decreases significantly, although not completely. At the same time there is a feature at 7.2 eV that becomes perceptible in the valence band XPS spectra obtained at temperatures above 600 °C.

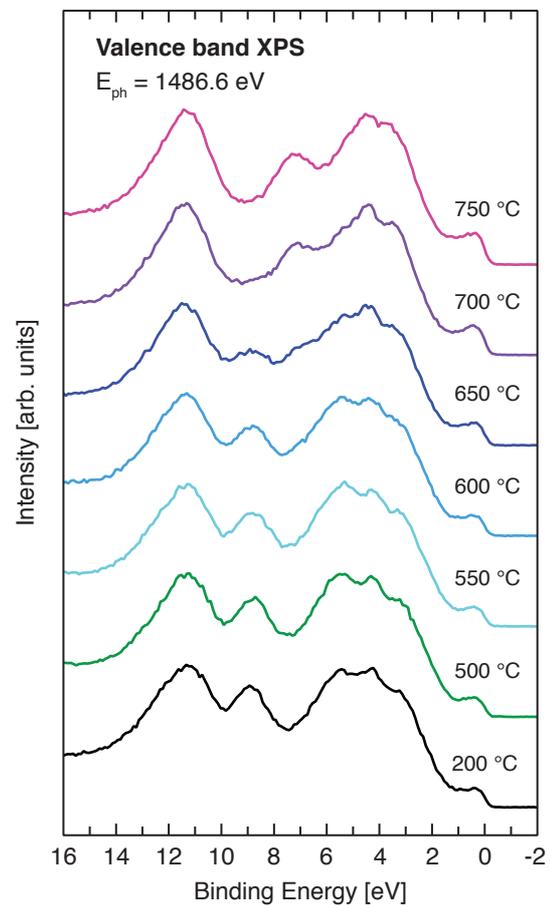

**Figure 3.** $Ti_3C_2T_x$ XPS spectra for the valence band region. The spectra of the $Ti_3C_2T_x$ sample heated to 200 and 500 °C show a very small difference only at 5.5 eV. Between 500 °C and 750 °C, in steps of 50 °C, there are significant changes in the binding energy region 4-10 eV. The photon energy was 1486.6 eV.

**Table 1.** Atomic subshell photoionization cross-sections [Mb] obtained from Yeh et al. [20]

| element | $E_{ph}$ [eV][a,b] | 1s | 2s | 2p | 3s | 3p | 3d | 4s |
|---|---|---|---|---|---|---|---|---|
| F | 1486.6 | 0.060 | 0.0028 | 0.00068 | | | | |
|   | 800.0  | 0.2923 | 0.013 | 0.0053 | | | | |
|   | 80.0   |       | 0.6742 | 3.508 | | | | |
|   | 40.8   |       | 0.5185 | 8.417 | | | | |
|   | 26.8   |       |        | 9.876 | | | | |
| O | 1486.6 | 0.040 | 0.0019 | 0.00024 | | | | |
|   | 800.0  | 0.2044 | 0.01 | 0.0025 | | | | |
|   | 80.0   |       | 0.6901 | 2.064 | | | | |
|   | 40.8   |       | 0.8342 | 6.816 | | | | |
|   | 26.8   |       |        | 9.772 | | | | |
| Ti | 1486.6 |      | 0.044 | 0.1069 | 0.0064 | 0.011 | 0.00017 | 0.00050 |
|    | 800.0  |      | 0.1289 | 0.5931 | 0.021 | 0.054 | 0.0017 | 0.0014 |
|    | 80.0   |      |       |        | 0.4308 | 1.090 | 1.472 | 0.067 |
|    | 40.8   |      |       |        |       |       | 4.012 | 0.1427 |
|    | 26.8   |      |       |        |       |       | 5.221 | 0.1851 |
| C | 1486.6 | 0.013 | 0.00066 | 1x10[-5] | | | | |
|   | 800.0  | 0.077 | 0.0033 | 0.00031 | | | | |
|   | 80.0   |       | 0.5440 | 0.3266 | | | | |
|   | 40.8   |       | 1.170 | 1.876 | | | | |
|   | 26.8   |       | 1.378 | 4.229 | | | | |
| Cl | 1486.6 |     | 0.022 | 0.031 | 0.0025 | 0.0019 | | |
|    | 800.0  |     | 0.078 | 0.195 | 0.0091 | 0.011 | | |
|    | 80.0   |     |       |       | 0.3493 | 0.9231 | | |
|    | 40.8   |     |       |       | 0.4020 | 0.6470 | | |
|    | 26.8   |     |       |       |       | 2.774 | | |

[a] 1486.6 corresponds to Al-Kα radiation.
[b] 40.8 corresponds to He-II radiation.

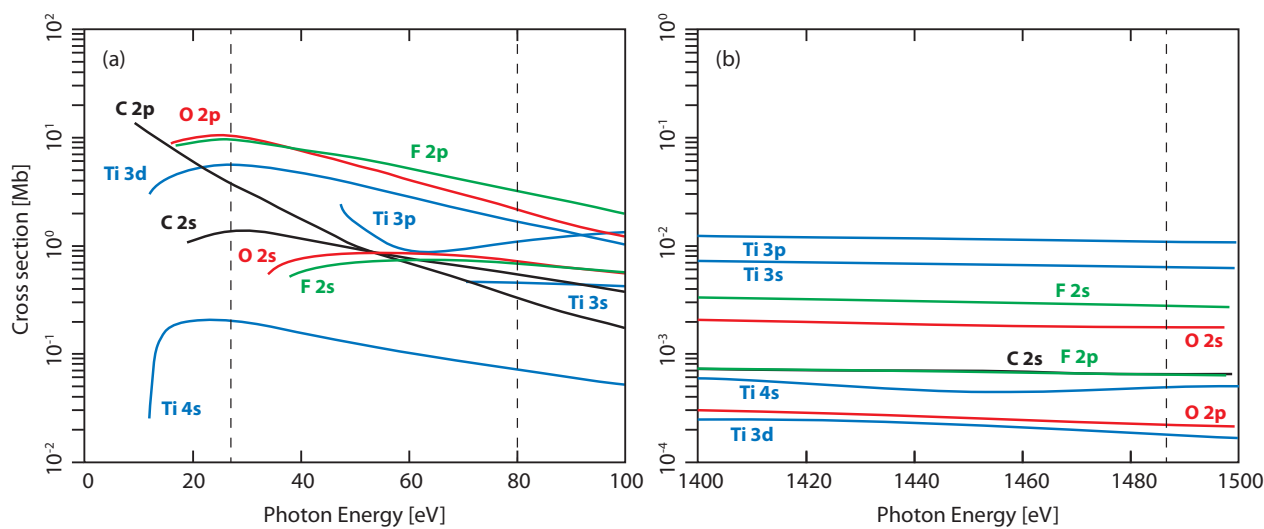

**Figure 4.** Calculated photoionization cross-sections for Ti, C, O, and F orbitals in (a) photon energy region 0-100 eV and (b) photon energy region 1400-1500 eV [20]. The dashed vertical lines in (a) indicate the photon energy range employed in the present valence band UPS study. The dashed line in (b) indicates the photon energy from Al-Kα radiation. Note the shift in the cross section by two-three orders of magnitude between the two regions.

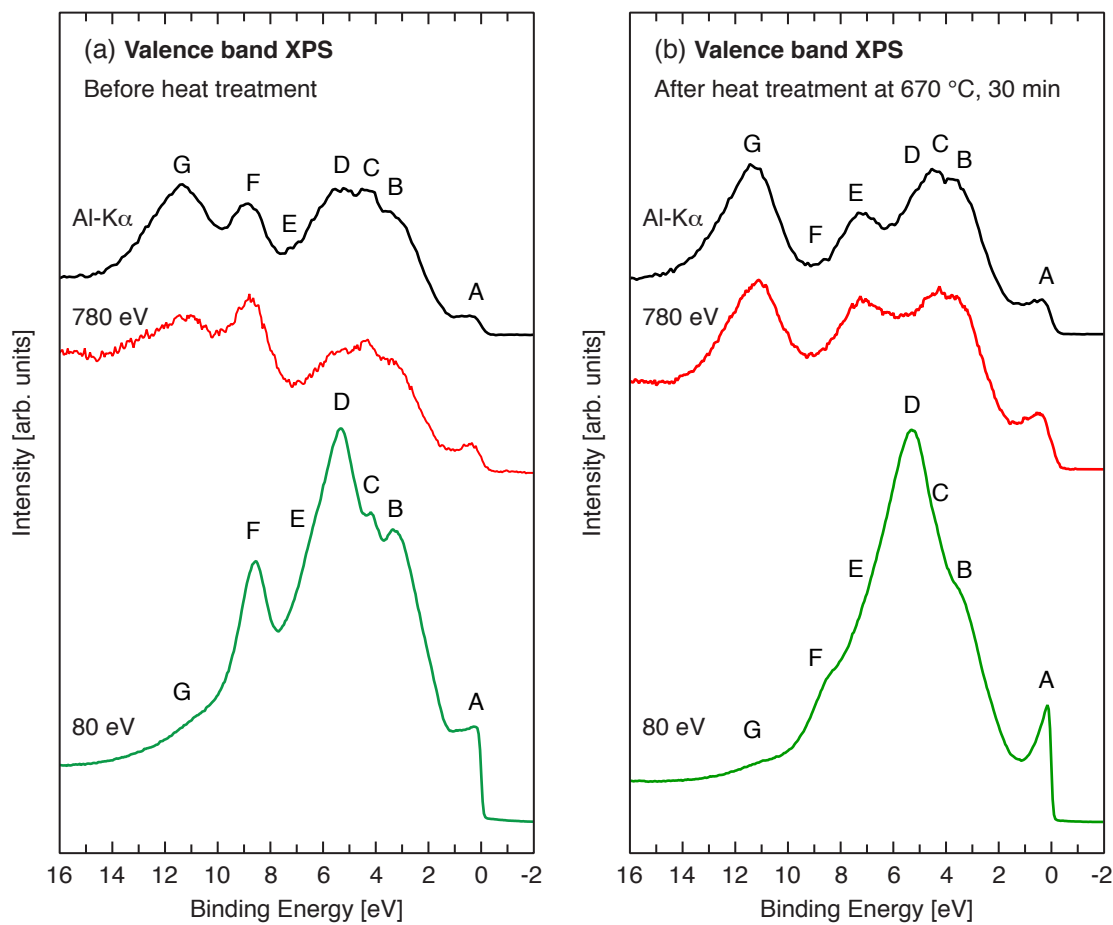

**Figure 5.** Valence band XPS spectra for Ti$_3$C$_2$T$_x$ (a) before and (b) after heat treatment at 670 °C, 30 min, using the photon energies 80 eV, 780 eV, and 1486.6 eV.

The temperature-programmed XPS presented in figures 2 and 3 show that the features observed at the center of the valence band originate from the termination species. The feature at 8.8 eV is suggested to be associated with F, which decreases when F desorbs as clearly proven by the F 1*s* XPS. The two features at 5.4 and 7.2 eV are proposed to be associated with O containing species where the latter suggest being O on the fcc-site as observed in the STEM study [14].

A valence band XPS spectrum of Ti$_3$C$_2$T$_x$ consists of photoelectron intensities of the higher occupied molecular orbitals, i.e., molecular orbitals that include mixing of Ti 3*p*, Ti 3*d*, C 2*s*, C 2*p*, O 2*p*, and/or F 2*p* orbitals. However, the obtained XPS intensity depends in part on the photoionization cross-sections of the probed elements [20]. Table 1 presents the calculated photoionization cross-sections for the selected orbitals of elements included in the valence band XPS spectra of Ti$_3$C$_2$T$_x$ at the photon energies 1486.6, 800, 80, 40.8, and 26.8 eV. According to the photoionization cross-sections there can be molecular orbitals that are observed in a valence band XPS spectrum of Ti$_3$C$_2$T$_x$ obtained using 1486.6 eV photon energy (Al-Kα radiation) that might be diminished in the corresponding valence band XPS spectrum obtained using, e.g., 80 eV photon energy. A well-known example is the corresponding valence band XPS spectrum of TiC that shows a dominating feature at 10.5 eV binding energy when the spectrum is obtained using 1486.6 eV photon energy but only a minor structure when the spectrum is obtained using 80 eV photon energy [30]. The loss in intensity of the feature at 10.5 eV binding energy for TiC was explained as a consequence of the different photoionization cross-section ratios between C 2*s*, Ti 3*p*, and Ti 3*d* for the two photon energies.

In figure 4 the calculated photoionization cross-sections for Ti, C, O, and F orbitals are presented in the photon energy regions 0-100 eV and 1400-1500 eV. In general, the photoionization cross-section is larger for the photon energy region 0-100 eV compared to 1400-1500 eV and it is, thus, an advantage to record valence band XPS in the lower photon energy region, which actually is in the range of UPS.

Figure 5 shows the valence band XPS spectra obtained using the photon energies 80, 780, and 1486.6 eV before and after heat treatment in the left and right panel, respectively.

All features are highlighted as A-G and they are all observed in at least one of the six spectra. The most apparent changes in the valence band XPS structures when different photon energies are employed are the large increase of intensity around 5.4 eV binding energy and the disappearing intensity around 11.3 eV when 80 eV photon energy is employed compared to when 780, and 1486.6 eV is used. The feature at 8.8 eV in the valence band XPS spectrum obtained using photon energy 1486.6 eV, shown in the figure 5(a), appears to shift to 8.6 eV when the spectrum was recorded using the photon energy 80 eV. That is, however, an illusion caused by the vanishing feature at 11.3 eV because when using photon energy of 1486.6 eV the high binding energy side of feature F is resting on the increasing slope of feature G but when photon energy of 80 eV is used the intensity of feature G decreases and subsequently also the intensity of the high binding energy side of feature F. Furthermore, the low binding energy side of feature F is resting on the decreasing slope of feature D, which has its intensity largely enhanced when the photon energy is 80 eV.

After the heat treatment at 670 °C for 30 min the valence band XPS spectra for the three selected photon energies 80, 780, and 1486.6 eV show less similarities. Feature A becomes sharper at lower photon energies while features B and C become less prominent. The center of the valence band XPS spectrum obtained using the photon energy 80 eV shows no resemblance with the corresponding spectra obtained using the photon energies 780 and 1486.6 eV. Surprisingly, feature E shows a prominent intensity at 7.2 eV when higher photon energies are employed but is not present in the valence band XPS spectrum obtained using the photon energy 80 eV. All these discrepancies make it difficult to assign the features in the center of the valence band XPS spectra in figure 5 to the termination species O, F, and OH without further analysis of the origin of the observed features. On the other hand are these discrepancies a fundamental necessity that will pave the way for a deeper understanding of the bond formation between the termination species and the $Ti_3C_2$-surface [19].

From figure 4 it is clear that one cause of the observed discrepancies between XPS spectra obtained at different photon energies presented in figure 5 is the variations in the photoionization cross-sections. O 2*p*, F 2*p*, and Ti 3*d* contributions dominate at the 80 eV photon energy while those orbitals have less influence at the photon energy 1486.6 eV. The valence band XPS spectra obtained using Al-K$\alpha$ radiation (1486.6 eV) are, on the contrary, dominated by Ti 3*p* contribution. That the feature at 11.3 eV is intense at the higher photon energies but disappears at the lower photon energies suggests that a photoionization cross-section around 1 MB at lower photon energies is too low in comparison with the dominant photoionization cross-sections of the O 2*p*, F 2*p*, and Ti 3*d* states, i.e., molecular orbitals that possess mainly C 2*s* and Ti 3*p* characters will not show features in the valence band XPS spectrum of $Ti_3C_2T_x$ at low photon energies, i.e., in the range of UPS. An analogous reasoning is valid for the disappearing feature at 7.2 eV, which is present when higher photon energies are employed but not when the photon energy 80 eV is used.

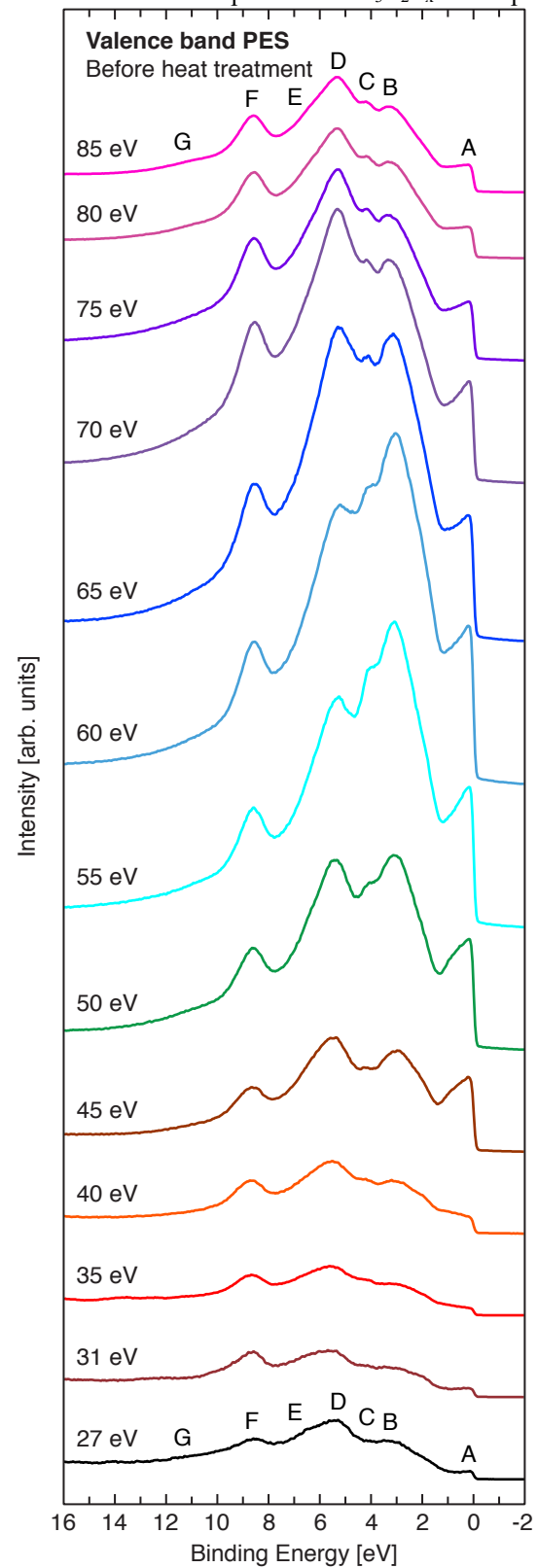

**Figure 6.** Resonant PES spectra in the valence band region for $Ti_3C_2T_x$ before heat treatment using the photon energies 27-85 eV.

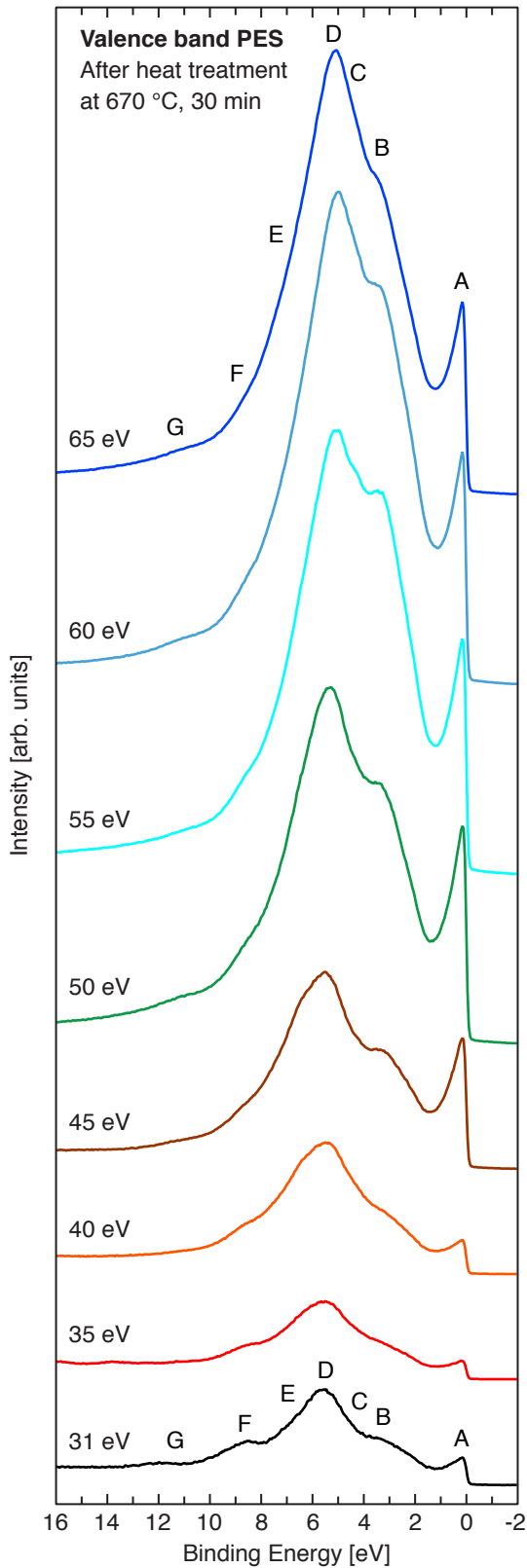

**Figure 7.** Resonant PES spectra in the valence band region for Ti$_3$C$_2$T$_x$ after heat treatment at 670 °C using the photon energies 31-65 eV.

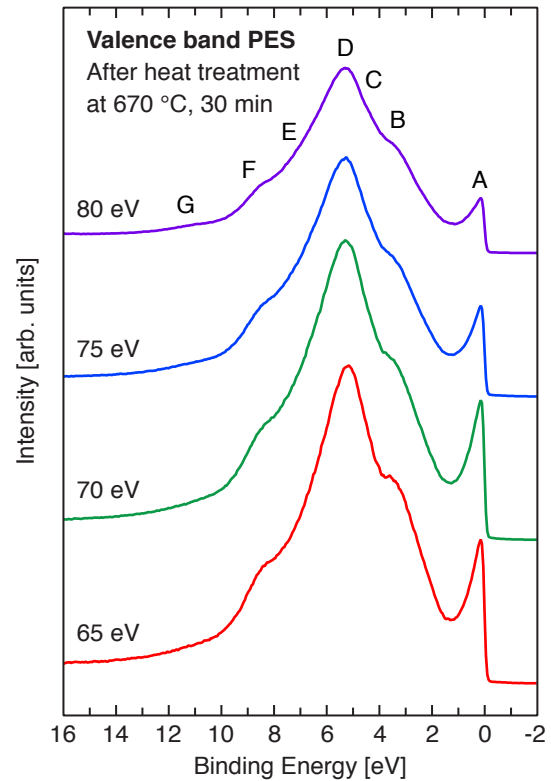

**Figure 8.** Resonant PES spectra in the valence band region for Ti$_3$C$_2$T$_x$ after heat treatment at 670 °C using the photon energies 65-80 eV.

Hence, when to study chemical bonding of the termination species in Ti$_3$C$_2$T$_x$ (T = F, O and/or OH) it is convenient to employ valence band UPS/XPS investigations using photon energies below 80 eV. At photon energies below 80 eV the photoionization cross-sections of the F 2*p* and O 2*p* orbitals are high and will provide significant intensities for features that originate from molecular orbitals with strong F 2*p* or O 2*p* character while the opposite is expected for molecular orbitals with strong C 2*s* or C 2*p* character.

### 3.3 Resonant photoemission spectroscopy

Figure 6 presents a series of valence band UPS/XPS spectra obtained using photon energies between 27 and 85 eV for the as-prepared freestanding Ti$_3$C$_2$T$_x$ film. Core-level XPS spectra (not shown) indicate that only a few O atoms occupy the fcc-site and that also Cl is present on the Ti$_3$C$_2$T$_x$ surface; the freestanding Ti$_3$C$_2$T$_x$ film was prepared using a solution of HCl(aq)+LiF(aq) as etchant. Although Cl is present on the as-prepared sample it desorbs completely already at a low temperature increase above room temperature, which suggests a relatively weak interaction with the $Ti_3C_2T_x$ surface. Hence, Cl is not a termination species, at least not for $Ti_3C_2T_x$ fully terminated by F and O. Further, the photoionization cross-section for Cl 3*p* in the low photon energy region is comparatively low and any intensity contribution in the valence band UPS/XPS spectrum would be insignificant, as shown in Table 1.

The first three spectra in figure 6, obtained using the photon energies 27, 31, and 35 eV, are quite similar and show the same features that were presented in a previous valence band UPS study where the photon energy was 21.21 eV (He I radiation) [17]. Features A, B, D, and F are clearly distinguishable while C, E, and G are not, which mainly is an effect of the high photoionization cross-sections of the F 2*p*, O 2*p*, and Ti 3*d* orbitals and the low photoionization cross-sections of the C 2*s*, C 2*p*, and Ti 2*p* orbitals. For the spectrum obtained at the photon energy 40 eV the total intensity starts to rise and with further increase in the photon energy some features enhance remarkably. The valence band UPS/XPS spectrum obtained at the photon energy 45 eV shows enhanced intensity, especially in the binding energy region 0-4 eV, i.e. in the Ti-C orbital hybridization region, where we have the features A, B, and C. The intensity in the binding energy region 0-4 eV dominates in the valence band UPS/XPS spectra obtained by the photon energies 50-60 eV but starts to decrease with further increases in photon energy. The valence band XPS spectra obtained using the photon energies 80 and 85 eV are close to identical.

Similar trends as in figure 6 are observed in figures 7 and 8, which show the analogous valence band UPS/XPS spectra obtained for $Ti_3C_2T_x$ after the heat treatment at 670 °C for 30 min. The corresponding core-level XPS spectra (not shown) show a decrease of the F content and a redistribution of O toward the fcc-site, i.e., same trend as shown in figure 2. The loss in F is manifested in the valence band UPS/XPS spectra as a large intensity decrease of feature F at 8.6 eV. Instead are the valence band UPS/XPS spectra dominated by the features A and D, which have significantly higher intensity in figures 7 and 8 compared to the valence band UPS/XPS measurements before the heat treatment shown in figure 6. The main reason is attributed to the removal of F in the heat treatment at 670 °C, which made it possible for O to fill the vacant fcc-sites [14] leading to a stronger feature D.

The reason why the series of valence band UPS/XPS spectra for the heat-treated $Ti_3C_2T_x$ are presented in two separated figures is because the obtained valence band UPS/XPS spectra are recorded at two different measurement spots. The removal of carbon components, such as graphite-like compounds, hydrocarbons, and alcohol contaminations, exposed the $Ti_3C_2T_x$ surface to more photons after the heat treatment compared to before the heat treatment when carbon contamination absorbed a significant amount of photons. The C 1*s* XPS obtained before and after the heat treatment showed the crucial removal of carbon contamination on the surface of the freestanding $Ti_3C_2T_x$ film. Without the photon absorbing carbon contamination the photon flux became more than the sample could withstand; features related to beam damages appeared after 7 hours of UPS/XPS measurements on the first measuring spot. Hence, the valence band UPS spectra at 65 eV photon energy in figures 7 and 8 were recorded at two different spots. Because of slightly different composition, surface roughness, and sample geometry there is an instant drop in intensity between the two spectra.

The valence band UPS/XPS spectra in figures 6-8 show intensity increase and decline as a function of photon energy. This behavior is because of resonant photoemission processes. The calculated photoionization cross-sections shown in figure 4(a) show that at the photon energies 27-80 eV the valence band features are dominated by O 2*p*, F 2*p*, and Ti 3*d* states. In addition, there is an onset of the photoionization cross-section of Ti 3*p* states in this photon energy region. A valence band UPS/XPS study of TiC found that the Ti 3*p* binding energy is around 34 eV [30], because the valence band UPS spectrum of TiC obtained at 34 eV photon energy produced a Ti 3*p* core-level peak close to 0 eV binding energy originated by the second-order reflection from the beamline monochromator. This implies that when approaching the Ti 3*p* x-ray absorption edge a 3*p* → 3*d* transition can occur, i.e., an electron from the Ti 3*p* orbital can excite to an unoccupied Ti 3*d* level. In the subsequent autoionization process an electron from the Ti 3*d* orbital decays into the Ti 3*p* hole while the excited electron in the Ti 3*d* orbital is ejected away. The process is called participant Auger-decay and produces the same final state as the one obtained in the direct Ti 3*d* photoemission process, i.e., a hole in the Ti 3*d* level and an escaping electron with the same kinetic energy as the Ti 3*d* photoelectron. Hence, the same initial and final states are involved in both the direct valence band photoemission and the autoionization process, which causes an interference and thus an intensity enhancement above the Ti 3*p* absorption edge for those features that have significant Ti 3*d* character [19,31-33].

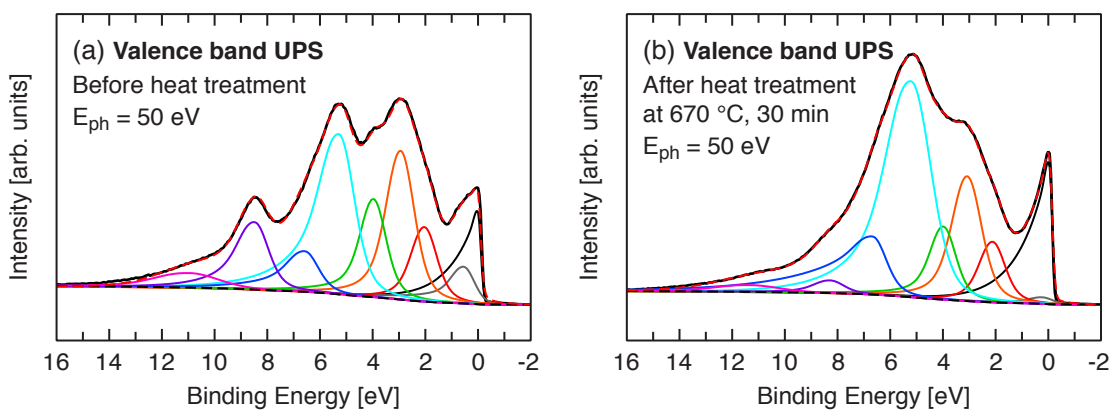

**Figure 9.** Curve fits of the valence band UPS spectra of $Ti_3C_2T_x$ obtained at the photon energy 50 eV (a) before and (b) after heat treatment at 670 °C. The black and the red dashed lines are the Shirley background and the accumulated intensity of all fitted components, respectively.

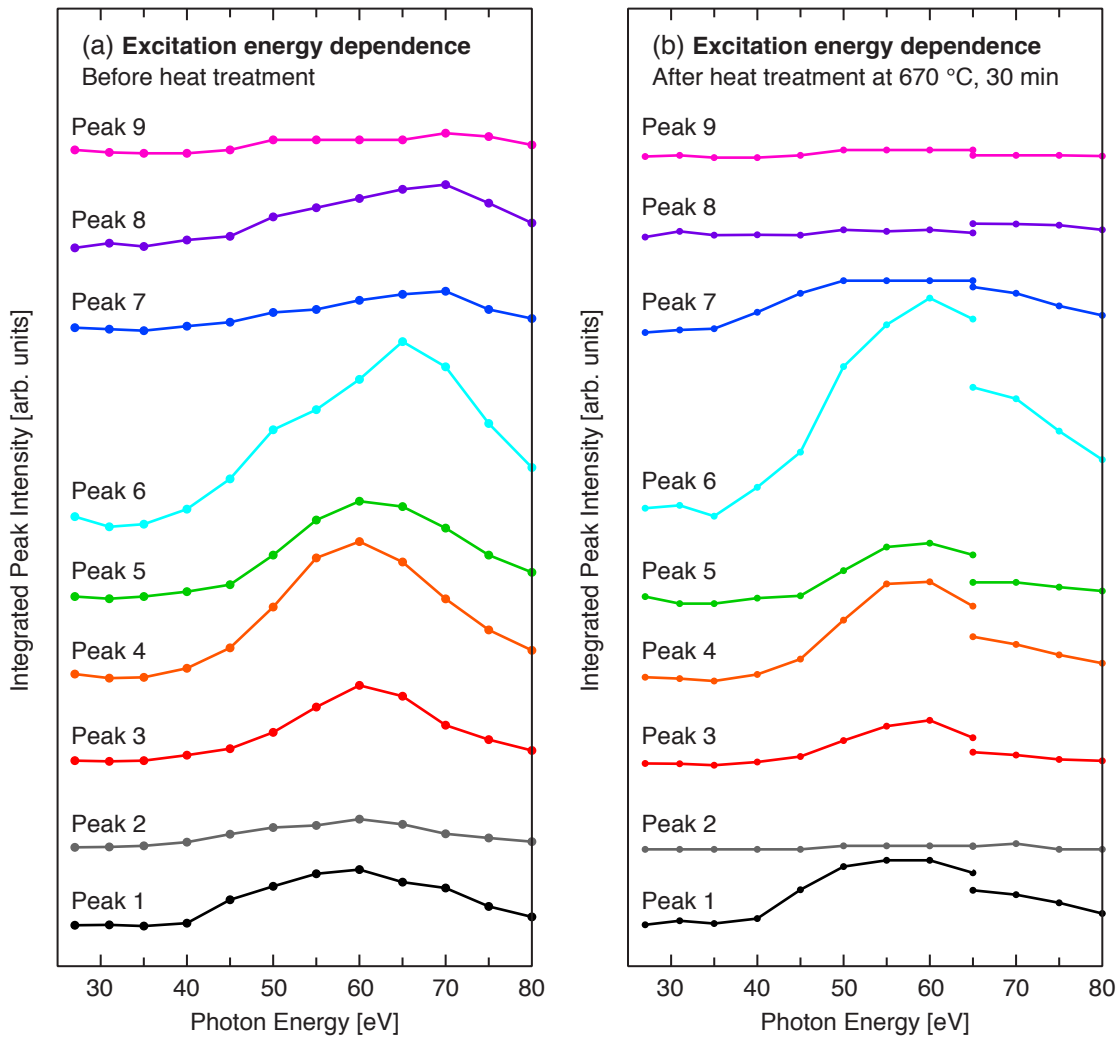

**Figure 10.** Integrated peak intensity as a function of photon energy for the nine peaks used in the curve fits of the valence band UPS/XPS spectra for $Ti_3C_2T_x$ (a) before and (b) after heat treatment at 670 °C obtained at the photon energies 27–80 eV.

To learn more about the components that contribute to the valence band UPS/XPS spectrum, and especially the origin of the features D, E, and F, the intensity distribution was fitted by asymmetric Gaussian-Lorentzian curves. The curves divide the valence band UPS/XPS spectrum into different segments and the purpose is to examine how each feature in the valence band UPS/XPS spectrum evolves with the photon energy. Two examples of curve fits are shown in figure 9 where the valence band UPS spectra obtained using 50 eV photon energy were curve fitted by nine asymmetric Gaussian-Lorentzian curves. The curve fittings of the valence band UPS/XPS spectra obtained before and after the heat treatment are shown in the left and right panels, respectively, of figure 9. Figure 10(a) shows how the nine curves change their integrated intensities as a function of photon energy for the curve fits of all valence band UPS/XPS spectra displayed in figure 6, i.e. before the heat treatment. The changes in the integrated intensities for all Gaussian-Lorentzian curves are insignificant at the lowest photon energies, but at 40 eV photon energy there is an onset where the integrated intensities begin to rise for all peaks, although very little for peaks 2, 7, and 9. Peaks 1-5 have their integrated intensity maxima at 60 eV photon energy while peak 6 has its maximum at 65 eV and peaks 7-9 have their maxima at 70 eV. That the peaks have their maximum integrated intensity at different photon energies is attributed to different bonding orbitals and their influence on the Ti 3*d*-states: peak 1-2 are attributed to Ti 2*p* – Ti 3*d* hybridization, peak 3-5 to C 2*p* – Ti 3*d* hybridization, peak 6 to O 2*p* – Ti 3*d* hybridization, and peak 8 to F 2*p* – Ti 3*d* hybridization. Peak 7 has contributions from both the O 2*p* – Ti 3*d* and the F 2*p* – Ti 3*d* hybridizations and peak 9 is assigned to C 2*s* – Ti 3*d* hybridizations. How much the integrated intensity increases depends on how strong interaction there is with the Ti 3*d* orbital. The enhancement of peak 4, which represents the C 2*p* hybridization with Ti orbitals, confirms the Ti 3*d* states involvement in the Ti-C bonds in $Ti_3C_2T_x$. Peaks 3 and 5 are parts of the C 2*p* – Ti 3*d* hybridization, although the higher intensity enhancement at photon energies above 60 eV for peak 5, compared to peaks 3 and 4, is because of the overlap with peak 6. The larger enhancement for peak 6 indicates a stronger interaction between O 2*p* and Ti 3*d* compared to the interaction between C 2*p* and Ti 3*d*, which suggests being the reason why O containing termination species cannot be removed from the $Ti_3C_2$-surface through heat treatments [14]. The small enhancement of peak 8, on the other hand, suggests a much weaker interaction between F 2*p* and Ti 3*d* compared to the Ti 3*d* interaction with both C 2*p* and O 2*p*. Peaks 2, 7, and 9 show only small intensity enhancements with photon energy, which mainly is because these peaks overlap with other peaks of considerable intensity increases. Hence, those three parts of the valence band UPS/XPS spectrum have minor contribution from Ti 3*d* orbitals.

Corresponding curve fittings were performed for the valence band UPS/XPS spectra obtained after the heat treatment at 670 °C for 30 min. The changes in the integrated intensities as a function of photon energy are shown in figure 10(b) for the curves fitted to the valence band UPS/XPS spectra displayed in figures 7-8. The trend of the integrated intensity enhancement after the 40 eV photon energy onset are the same as before the heat treatment, except for peaks 2, 8, and 9. That peaks 8, and 9 do not show any variations in the integrated intensities with the photon energy is because most of the F on the $Ti_3C_2$-surface is removed in the heat treatment and the integrated intensities for peak 8 and 9 are therefore relatively small and variations in the integrated intensities are, thus, too small to be perceived. The discontinuity in integrated intensity for all peaks at 65 eV photon energy in figure 10(b) is because of the two measurement spots used; the first measurement spot showed indications of beam damage after the series of valence band UPS/XPS spectra obtained with the photon energies 34-65 eV. An interesting difference in the amplitude enhancement of peak 6 obtained for the two curve fittings of the valence band UPS/ XPS spectra before- and after the heat treatment is the shift of amplitude maximum to 60 eV after the heat treatment, which suggest being an effect of the O filling the fcc-sites that become available when F atoms have been removed [14].

### 3.4 Bond formation in $Ti_3C_2T_x$

On account of the changes in the photoionization cross-sections and, in addition, the resonant Ti 3*p* x-ray absorption with the subsequent participant autoionization process it is possible to acquire knowledge about the Ti-C bond formation as well as the bond formation between the termination species and the $Ti_3C_2$-surface.

The pronounced intensity in the binding energy region 2-4 eV in the valence band XPS spectra obtained at both 80 and 1486.6 eV photon energies, where the spectra obtained at 80 eV photon energy is dominated by Ti 3*d* states and the spectra obtained at 1486.6 eV photon energy is dominated by Ti 3*p* states, suggest that the bond formation between Ti and C includes both Ti 3*p* and Ti 3*d* orbitals. The resonant behavior around the Ti 3*p* x-ray absorption edge further suggests a dominant Ti 3*d* contribution in the 2-4 eV binding energy region. Hence, the rigid Ti-C bonding in $Ti_3C_2T_x$ is a result of the strong orbital hybridization between the Ti 3*p*, Ti 3*d*, and C 2*p* orbitals, which is similar as in cubic TiC [30]. In addition, the Ti 3*p*-states are involved in an interaction with the C 2*s* orbital, which can be concluded from the feature at 11.3 eV in the spectra obtained at higher photon energies that is not present in the spectra obtained at lower photon energies. Similar as for TiC it cannot be excluded that Ti 4*p* is involved in the Ti-C bond formation also in $Ti_3C_2T_x$.

As in the interaction between Ti and C, the bondings between the O containing termination species and the $Ti_3C_2$-surface are results of the strong orbital hybridization between the Ti 3*p*, Ti 3*d*, and the O 2*p* orbitals. The Ti 3*d* – O 2*p* orbital hybridization dominates the valence band UPS/XPS spectra at low photon energies with an intensity maximum at 5.3 eV binding energy both before and after the heat treatment, even though the O 1s XPS in figure 2 shows intensity redistribution indicating a change in the O containing termination species. However, as with the Ti-C feature at 11.3 eV, which shows a profound intensity at higher photon energies but almost no intensity at lower photon energies, there is a feature at 7.2 eV in the valence band XPS spectra of the heat treated $Ti_3C_2T_x$ obtained at higher photon energies that can be assigned to O on the fcc-site that shows no intensity at lower photon energies. The resemblance with the Ti-C feature at 11.3 eV suggests that the 7.2 eV feature in the valence band XPS spectra obtained at higher photon energies originates from an interaction between O and Ti orbitals providing a hydridized orbital with relatively high photoionization cross-section at higher photon energies. In addition, the involved Ti orbital must have a relatively low photoionization cross-section ($\leq 1$ Mb) when lower photon energies are employed. An electron orbital that fulfill these criteria is the Ti 4$p$ orbital. The Ti 4$p$ orbital is not occupied in ground state conditions of Ti and is therefore excluded in figure 4. Nevertheless, the photoionization cross-section for Ti 4$p$ would be in between those of Ti 3$s$ and Ti 4$s$, see for example the photoionization cross-sections for Zr in [20], which would be a too low photoionization cross-section at lower photon energies but significant at higher photon energies. Earlier studies of TiC have suggested that Ti 4$p$ hybridizes with C 2$s$ and 2$p$ forming filled 1t$_{1u}$ and 2t$_{1u}$ levels, respectively [30], and it is therefore not a far-fetched suggestion that Ti 4$p$ might be involved in bonding with O on the fcc-site. The recent Ti K-edge XAS study found that the main-edge features that originate from the Ti 1$s$ → 4$p$ excitation showed sensitivity toward termination species on the fcc-site [21]. Hence, the strong O bonding toward the Ti atoms at the fcc-site involves the O 2$p$, Ti 3$p$, Ti 3$d$, and Ti 4$p$ orbitals.

The feature at 8.6 eV can be assigned to F atoms occupying the fcc-site, because it diminishes simultaneously with the decrease of the F 1$s$ XPS intensity during the heat treatment. The F feature remains clearly visible at both higher and lower photon energies, which suggests that the intensity contribution mainly consists of F 2$p$-states. That the F feature does not show a significant enhancement when the photon energy is close to the Ti 3$p$ x-ray absorption edge, i.e. a small resonance effect, suggests that Ti 3$d$ orbitals are not significantly involved in the bonding with the F atoms. Hence, the bonding between the termination species F and the Ti$_3$C$_2$-surface is mainly a result of an orbital hybridization between the F 2$p$ and Ti 3$p$ orbitals.

The last part of the Ti$_3$C$_2$T$_x$ valence band XPS spectrum to scrutinize is the region near E$_f$. The resonant intensity enhancement when the photon energy is close to the Ti 3$p$ x-ray absorption edge indicates a predominant Ti 3$d$ contribution near E$_f$. Another interesting observation is the intensity rearrangement from the two features before heat treatment to one sharp feature after the heat treatment. The comparison between the valence band UPS/XPS spectra of Ti$_3$C$_2$T$_x$ before and after heat treatment shows a resemblance with the corresponding comparison between TiC and TiN [30]. Before the heat treatment the Ti$_3$C$_2$T$_x$ valence band UPS/XPS spectra appears to be a mixture of TiC-like and TiN-like compounds and after the heat treatment the Ti$_3$C$_2$T$_x$ valence band XPS spectra dominates by a TiN-like compound that have some minor contribution of a TiC-like compound. In other words, the O atoms that occupy the fcc-sites in Ti$_3$C$_2$T$_x$ contribute with electrons to molecular orbitals with Ti 3$d$ character similarly as the N atoms contribute with electrons to the corresponding molecular orbitals in TiN. The F atoms occupying the fcc-sites do not.

### 3.5 Computed valence band UPS/XPS spectra

In order to obtain more information about the interaction between the termination species and the Ti$_3$C$_2$-surface, valence band UPS/XPS spectra were computed based on DOS calculations of Ti$_3$C$_2$T$_x$ (T = F, O, and OH) weighted by dipole transition matrix elements and photoionization cross-sections. Figure 11 presents computed valence band UPS/XPS spectra of a Ti$_3$C$_2$-layer without termination species. The figure shows that the Ti 3$d$-states dominate the valence band UPS spectra, i.e., the spectra obtained using low photon energies, while the Ti 3$p$-states dominate when the valence band XPS spectrum is computed using Al-K$\alpha$ radiation. The variation in the dominating contribution follows the calculated photoionization cross-sections shown in figure 4. Other noteworthy findings are the enhanced C 2$s$ states and an increased intensity around 5 eV when using Al-K$\alpha$ radiation compared to the valence band UPS spectra

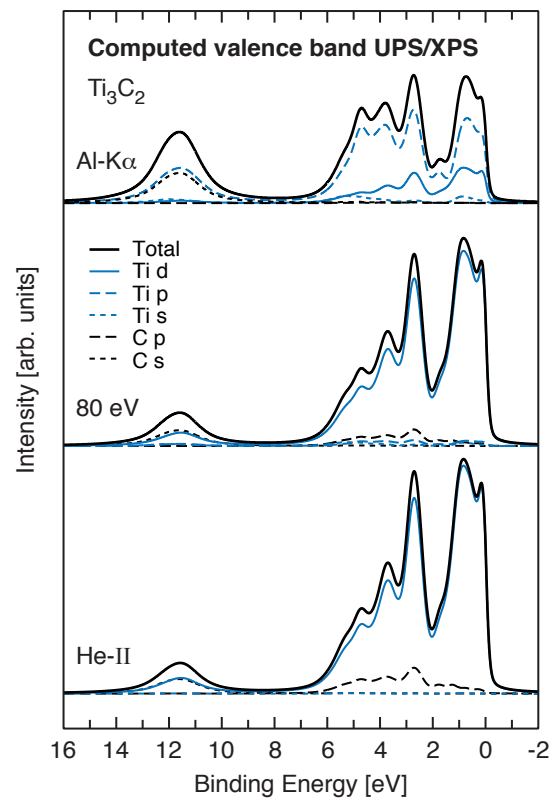

**Figure 11.** Computed valence band UPS/XPS spectra for Ti$_3$C$_2$ without terminating species for the photon energies 40.8 eV (He-II radiation), 80 eV, and 1486.6 eV (Al-K$\alpha$ radiation).

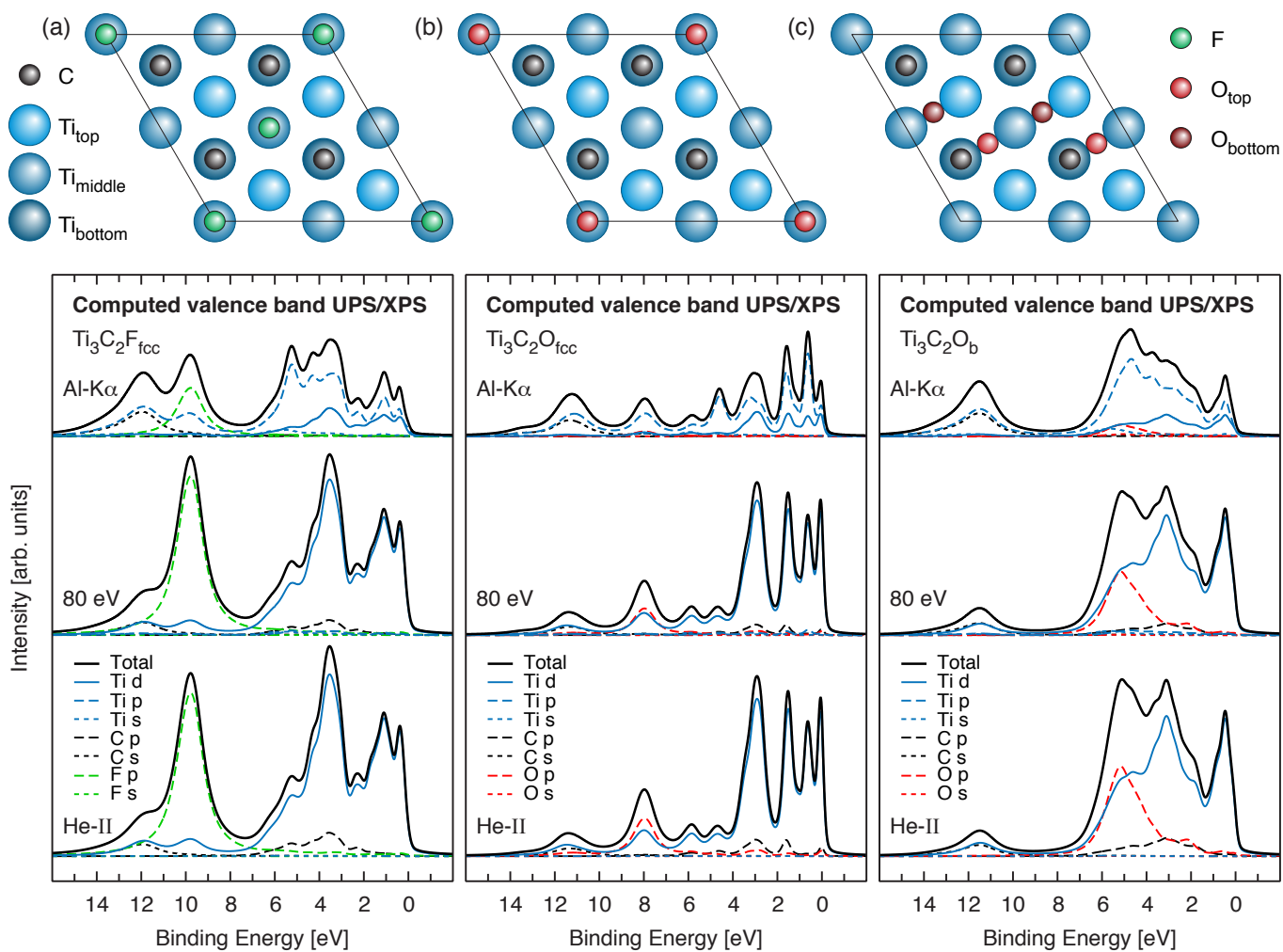

**Figure 12.** Computed valence band UPS/XPS spectra obtained from the crystal structure of (a) $Ti_3C_2F_{fcc}$ where F sits on fcc-sites, (b) $Ti_3C_2O_{fcc}$ where O sits on fcc-sites, and (c) $Ti_3C_2O_b$ where O sits on bridge sites. The three valence band UPS/XPS spectra in the panel below each crystal structure are computed for the photon energies 40.8 eV (He-II radiation), 80 eV, and 1486.6 eV (Al-K$\alpha$ radiation). Darker Ti and O atoms in the crystal structures are in the middle and bottom monolayers.

using 40.8 (He-II radiation) and 80 eV photon energy.

Considering the simple model used in the study the computed valence band UPS/XPS of the bare $Ti_3C_2$-layer correlate to a certain extent with the Ti-C features A-C, and G in the experimentally recorded valence band UPS/XPS spectra shown in figures 5-8, except the enhanced intensity at 0.8 eV in the computed spectra. Hence, the Ti-C distances obtained from the recent Ti K-edge XAS study [22] are appropriate in the $Ti_3C_2T_x$ structure model.

Figure 12 shows the computed valence band UPS/XPS spectra where the simple model calculations included the interactions between the $Ti_3C_2$-surface and the termination species F on the fcc-site ($F_{fcc}$), O on the fcc-site ($O_{fcc}$), and O on the bridge site ($O_b$), see figure 1 and the structure models above the UPS/XPS spectra panels in figure 12(a), (b), and (c), respectively. The computed valence band UPS/XPS spectra of $Ti_3C_2F_{fcc}$ shows two distinct changes compared with the corresponding computed valence band UPS/XPS of the termination-free $Ti_3C_2$-layer: the F 2*p* feature at 9.8 eV and the shift toward higher binding energy for the enhanced Ti 3*d*-states at 3.5 eV. The interactions between the $Ti_3C_2$-surface and the termination species $O_{fcc}$ and $O_b$ show also distinct changes in the computed valence band UPS/XPS spectra. The computed spectra of $T_3C_2O_{fcc}$ show several narrow features distributed over the valence band region while the computed spectra of $T_3C_2O_b$ show broad structures in the 2-6 eV binding energy range. Hence, the computed spectra presented in figure 12 show how sentitive the valence band features are for different termination species and selected bonding configuration.

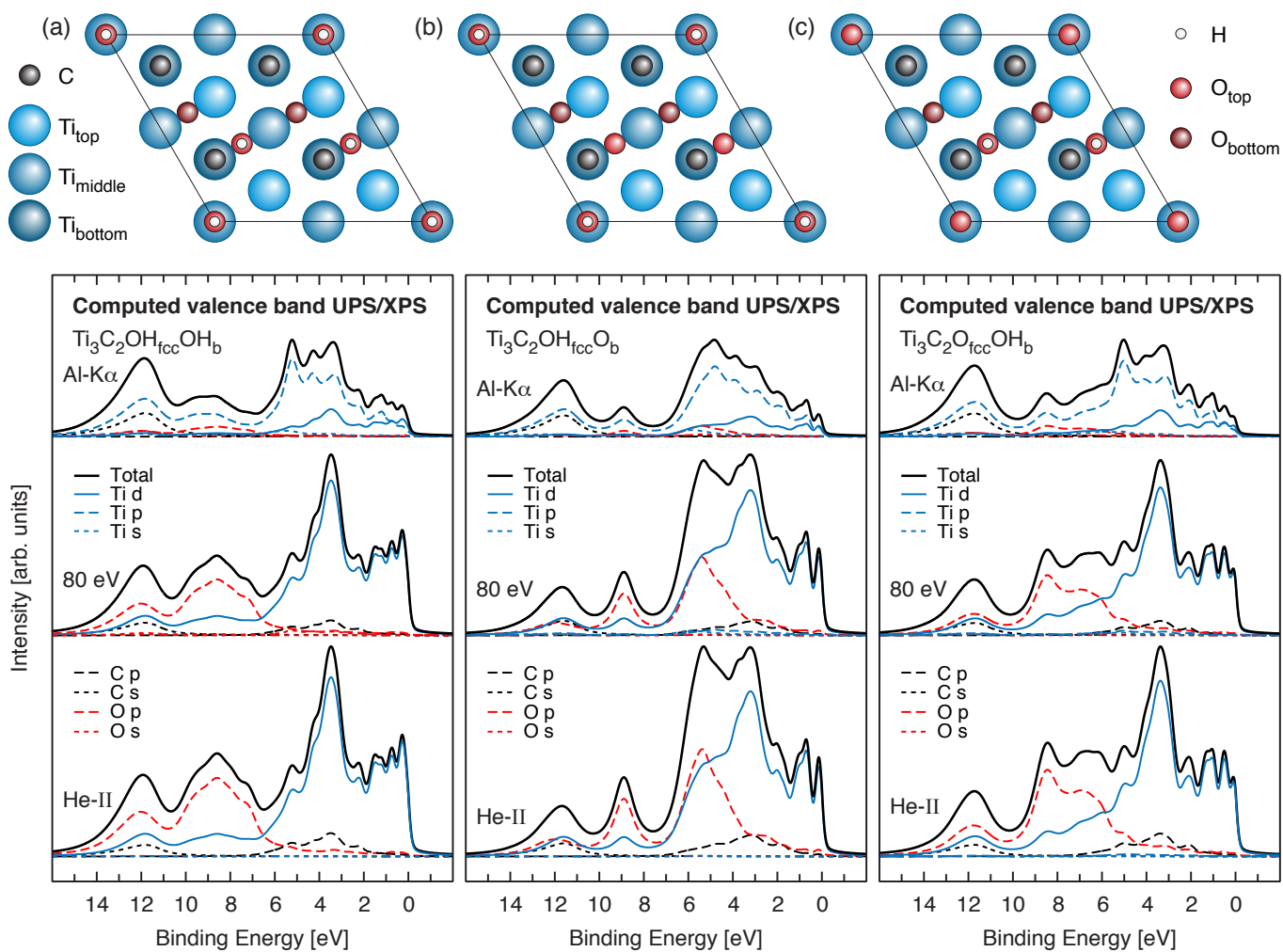

**Figure 13.** Computed valence band UPS/XPS spectra obtained from the crystal structure of (a) $Ti_3C_2OH_{fcc}OH_b$ where OH sits on both fcc- and bridge sites, (b) $Ti_3C_2OH_{fcc}O_b$ where OH sits on fcc-sites and O on bridge sites, and (c) $Ti_3C_2O_{fcc}OH_b$ where OH sits on bridge sites and O on fcc-sites. The three valence band UPS/XPS spectra in the panel below each crystal structure are computed for the photon energies 40.8 eV (He-II radiation), 80 eV, and 1486.6 eV (Al-Kα radiation). Darker Ti and O atoms in the crystal structures are in the middle and bottom monolayers.

The core-level XPS spectra in figure 2 show that there is not only one O-containing termination species on the $Ti_3C_2$-surface. The most popular suggestion is OH and sometimes together with O on different sites. Figure 13 presents the computed valence band UPS/XPS spectra for mixed termination models of $Ti_3C_2OH_{fcc}OH_b$, $Ti_3C_2OH_{fcc}O_b$, and $Ti_3C_2O_{fcc}OH_b$, i.e. where OH sits on both the fcc-site and the bridge site, on the fcc-site together with O on the bridge site, and on the bridge site together with O on the fcc-site, respectively. The computed spectra obtained using 1486.6 eV photon energy (Al-Kα radiation) show significant intensity around 8.5 eV binding energy that originates from OH on the $Ti_3C_2$-surface and this feature shows an intensity increase when lower photon energies are employed. In addition there is a second feature around 12 eV binding energy with significant intensity at lower photon energies. The two OH-derived features are, thus, clear signs of OH-terminations if that would be an actuality. A comparison with the corresponding experimental valence band UPS/XPS spectra in Figures 7 and 8 reveal that the OH-derived features are absent in the experimental spectra and, thus, makes the assumption of OH as a termination species in $Ti_3C_2T_x$ unrealistic. At least not at a quantity that is noteworthy.

Figure 14 shows, on the other hand, the computed valence band UPS/XPS spectra where the model calculations included the combined interactions between the $Ti_3C_2$-surface and the termination species $F_{fcc}+O_{fcc}+O_b$ and $O_{fcc}+O_b$ as shown in the structures models above the UPS/XPS spectra panels in figure 14(a) and (b), respectively. The agreement with the experimentally recorded valence band UPS/XPS spectra for $Ti_3C_2T_x$ before and after heat treatment shown in Figures 5-8 is fairly good for both the $Ti_3C_2(F_{fcc}+O_{fcc}+O_b)$ and the $Ti_3C_2(O_{fcc}+O_b)$ model computations, although the amount of O is too low in the models and therefore do not provide enough intensity around 5 eV. Hence, the comparison between the DFT computed model in figure 14(a) and the experimentally obtained valence band UPS/XPS spectra of $Ti_3C_2T_x$ before the heat treatment presented in the figure 5(a) reaveals that the inherently formed termination species on $Ti_3C_2T_x$ are F and O occupying the fcc-site and that O also occupies a bridge site between two surface Ti atoms. Furthermore, when including figure 13 in the comparison it is clear that the most popular suggestion OH is not an inherently formed termination species on $Ti_3C_2T_x$.

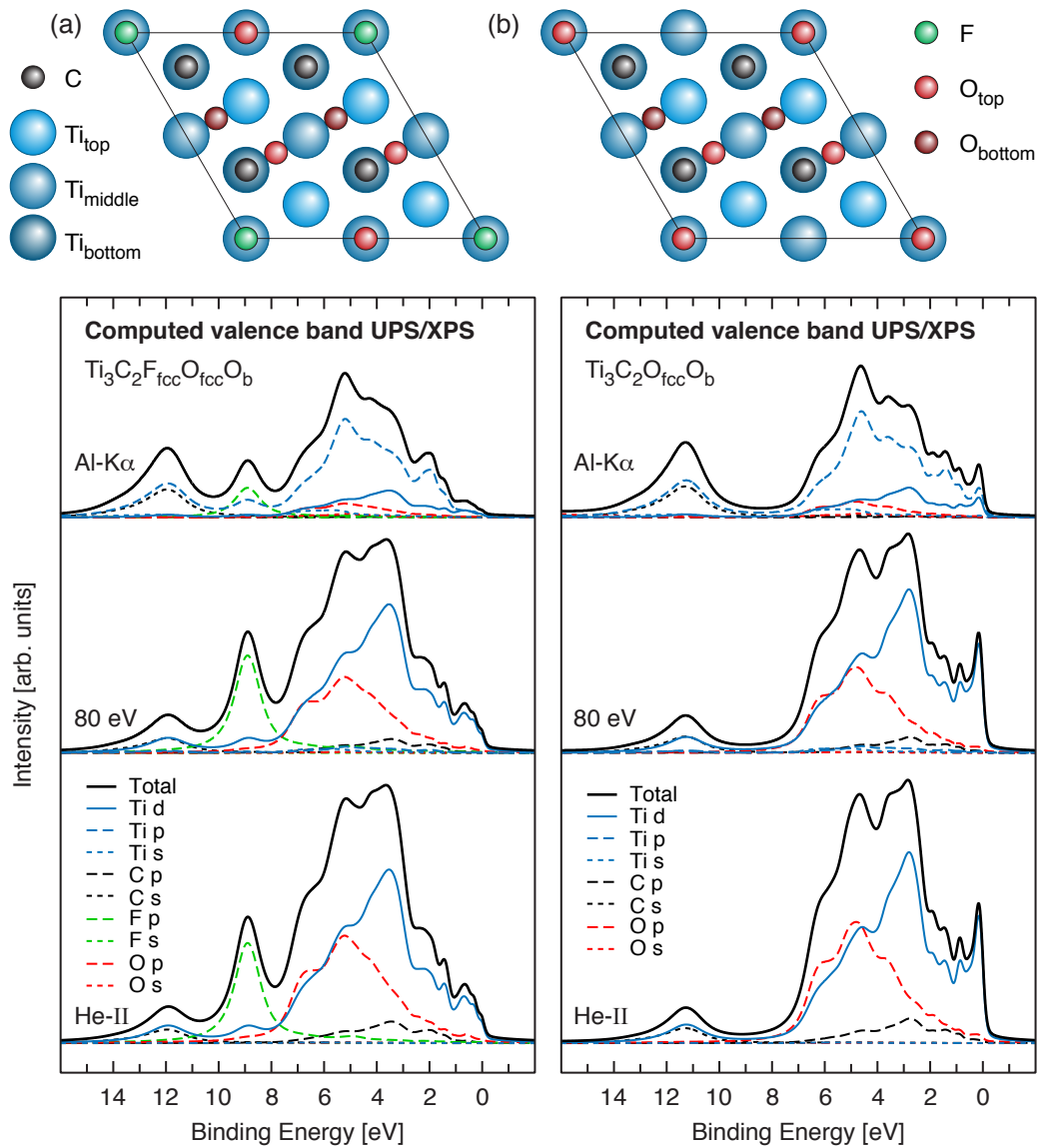

**Figure 14.** Computed valence band UPS/XPS spectra obtained from the crystal structure of (a) $Ti_3C_2F_{fcc}O_{fcc}O_b$ where F sits on fcc-sites and O sits on both fcc- and bridge sites and (b) $Ti_3C_2O_{fcc}O_b$ where O sits on both fcc- and bridge sites. The three valence UPS/XPS band spectra in the panel below each crystal structure are computed for the photon energies 40.8 eV (He-II radiation), 80 eV, and 1486.6 eV (Al-Kα radiation). Darker Ti and O atoms in the crystal structures are in the middle and bottom monolayers.

Equally importantly is the realization of including O on the bridge site between two Ti atoms in the $Ti_3C_2T_x$ structure model. The result shows that a significant amount of $O_b$ is present both before and after the heat treatment. Because the total amount of $O_b+O_{fcc}$ remains unchanged during a heating process [14] the amount of $O_b$ after the heat treatment depends on the $O_b/F_{fcc}$ ratio and the amount of $F_{fcc}$ that has desorbed. When the ratio between $O_b$ and $F_{fcc}$ is high the distribution of $O_b$ will cover the whole $Ti_3C_2$-surface even after all $F_{fcc}$ atoms have been removed. The valence band UPS/XPS spectrum of $Ti_3C_2O_x$ using low photon energies will, thus, display the features shown in figures 7-8 and the computed $Ti_3C_2(O_{fcc}+O_b)$ spectra in figure 14(b). However, if the $O_b/F_{fcc}$ ratio is low it is plausible that domains with only $O_{fcc}$ are formed on the $Ti_3C_2$-surface, which then would provide a valence band UPS/XPS spectrum similar to the computed $Ti_3C_2(O_{fcc}+O_b)$ spectrum in figure 14(b) but with an extra feature around 8 eV as in the $Ti_3C_2O_{fcc}$ model computation shown in figure 12(b).

### 3.6 Bonding configuration of termination species

The distance between F and the surface Ti atoms as well as the distance between O on the bridge site to the surface Ti atoms used in the $Ti_3C_2T_x$ structure model (2.115 and 1.979 Å, respectively) are close to the obtained distances in the recent Ti K-edge XAS study [21]. The distance between O on the fcc-site and the surface Ti atoms (1.961 Å) is, on the other hand, somewhat longer. The local configuration for $O_{fcc}$ in the $Ti_3C_2T_x$ structure model includes the $O_{fcc}$ in the center of a tetrahedron bonded to three surface Ti atoms positioned in the vertices of the tetrahedron base plane as shown in figure 15. The Ti-O-Ti bond angle, i.e., the angle between lines from the $O_{fcc}$ in the center of the tetrahedron to any two Ti atoms, is 105.3°, which is not far from the corresponding angle between the accepting H-bonds in the solvation structure of OH$^-$(aq) [34-37]. It has been proposed that OH is the termination species in the three-fold hollow fcc-site instead of O [15,16], although with OH in the center of a tetrahedron bonded to three surface Ti atoms positioned in the vertices of the tetrahedron base plane the H would be in the direction toward the vertex opposite the base plane. The analogy between the OH on the fcc-site and an OH$^-$ in aqueous solution implies strong bonds to the Ti surface atoms but a very weak bond to intercalated species [36,37]. Hence, OH on the fcc-site would have serious implications in intercalation processes, especially toward cations important in, e.g., energy storage applications [38]. An O on the fcc-site will, on the other hand, have an electron lone pair lobe, i.e. a charge cloud, that extends out from the $O_{fcc}$ away from the $Ti_3C_2$-surface in the direction toward the vertex opposite the base plane acting as an "anchor point" for intercalated species.

The F on the fcc-site shows a Ti-F-Ti bond angle of 94.0°, which is smaller compared to the corresponding Ti-O-Ti bond angle of $O_{fcc}$. The smaller bond angle is a consequence of the weaker bonding between F and the three surface Ti atoms that form the fcc-site, which is derived from the additional electron, compared to O, in the F $2p$ orbitals. Hence, the driving force to form bonding orbitals with the Ti $3d$ is largely reduced.

That O bonds to two Ti surface atoms in a bridge site is not an entirely unpredictable conclusion. The local configuration for $O_b$ in the $Ti_3C_2T_x$ structure model, with a Ti-O-Ti angle of 103.9°, is similar to the tri-atomic construction of $H_2O$ where one oxygen atom (O) is bonded to two hydrogen atoms (H) with an H-O-H angle of 104.5° [39]. That the Ti-O-Ti angle is slightly less than the H-O-H angle in the water molecule indicates a smaller bond pair repulsion, i.e., that the bonding electrons are somewhat farther away from the O when bonded to two Ti compared to when bonded to two H. In addition, the analogy between the O on the bridge site and an $H_2O$ molecule includes the two electron lone pair lobes that extend out from the $O_b$ away from the $Ti_3C_2$-surface. These two electron lone pair lobes might be "anchor points" for intercalation processes and, thus, play important roles in, e.g., energy storage applications.

Both the experiments and the computations show that the valence band UPS/XPS features in the binding energy region 0-4 eV are dominated by contributions from the Ti $3d$ character in the molecular orbitals close to $E_f$ and contribution from the Ti $3d$ – C $2p$ hybridization around 3 eV below $E_f$. That the intensity will increase substantially because of the interference between the direct photoelectron process and the participant autoionization process suggests that Ti $3p$ is mixed into the Ti $3d$ orbitals near $E_f$ and into the Ti $3d$ – C $2p$ hybridization around 3 eV. The O contribution to the valence band UPS/XPS spectra around 5 eV binding energy is dominated by Ti $3p$ – O $2p$ hybridization with significant Ti $3d$ contribution. The features originating from O on the fcc-site are superposed on the features from O on the bridge site between two surface Ti atoms. A feature at 7.2 eV in the valence band XPS spectra obtained using higher photon energies suggests that also Ti $4p$ states are involved in the bonding between Ti and O on the fcc-site. The bonding between Ti and F on the fcc-site is, on the other hand, mainly through Ti $3p$ – F $2p$ hybridization, which manifests as a clear peak at 8.6 eV in the valence band UPS/XPS spectra of which the amplitude depends mainly on the surface coverage and the photoionization cross-section.

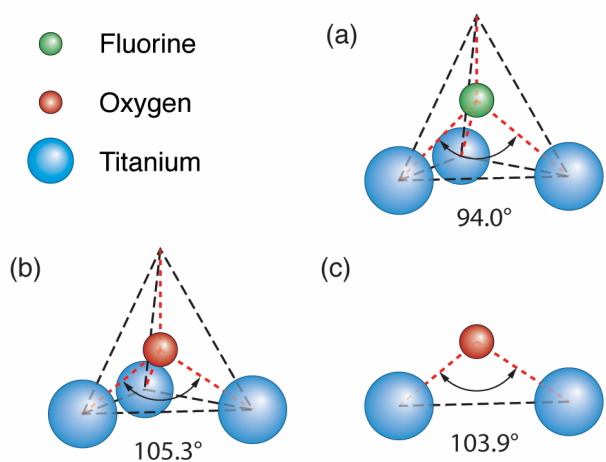

**Figure 15.** The Ti-F/O-Ti bond angle for (a) F on the fcc-site, (b) O on the fcc-site, and (c) O on the bridge site. Larger bond angle corresponds to stronger bonds toward the Ti atoms.

The computed valence band UPS/XPS spectra show that OH as termination species would add additional features with significant intensity in the binding energy region of 8-14 eV below $E_f$, which the experimentally recorded valence band UPS/XPS spectra do not display. The assumption of OH as a termination species is thus refuted.

The conclusion from the valence band UPS/XPS study that OH is not an inherently formed termination species in $Ti_3C_2T_x$ might come as a surprise but does not contradict other studies. A combined temperature-programmed XPS/STEM study found no indications of that OH would be a termination species [14] and previous XPS studies have just assumed that the broad feature at 531.8 eV in the O 1*s* XPS spectrum includes an OH-component [15] although it was not motivated by the shape of the spectrum; the combined temperature-programmed XPS/STEM study showed that the broad feature at 531.8 eV in the O 1*s* XPS spectrum originates from O on the fcc-site. However, a nuclear magnetic resonance (NMR) study of $Ti_3C_2T_x$ powders found a signal that was assigned to OH [16]. The NMR-study could not provide evidences of that the detected OH was bonded to the $Ti_3C_2$-surface, although it was found that the OH was located not far away from C. Even so, $Ti_3C_2T_x$ is prone to oxidation [22] and it is established that OH bonds strongly to $TiO_2$ surfaces with O-vacancy sites [40-46], which argues that the obtained OH signal rather originates from OH adsorbed on oxidized material. If so, the OH would still be in close proximity of C because $Ti_3C_2T_x$ is a 2D material and oxidized sections would therefore be only a few Ångström thick. Hence, the suggestion of OH as an inherently formed termination species in $Ti_3C_2T_x$ is merely based on an assumption, but even though it is a reasonable assumption the present study clearly shows that OH-features are missing in the experimental valence band UPS/XPS spectra of $Ti_3C_2T_x$ and that the $Ti_3C_2$-surface instead is inherently terminated by only F and O.

Table 2 summarizes the observed features, the assignments, and the involved molecular orbitals based on the resonant PES investigation in the valence band energy region using the photon energies 27-80 eV as well as the valence band UPS/XPS spectra obtained at 80, 780, and 1486.6 eV photon energies. The table lists the dominant contributions to the valence band UPS/XPS spectra although hybridization with Ti 4*p* cannot be excluded at multiple features.

**Table 2.** The binding energy (BE) and assignment of the features in the valence band UPS/XPS spectrum of the as prepared $Ti_3C_2T_x$.

| feature notation | BE [eV] | assignment[a,b,c] | dominating contributions |
|---|---|---|---|
| A | 0.2 | Ti | Ti 3*d* + Ti 3*p* |
| B | 3.2 | Ti-C | Ti 3*d* + Ti 3*p* + C 2*p* |
| C | 4.2 | Ti-C | Ti 3*d* + Ti 3*p* + C 2*p* |
| D | 5.3 | Ti-$O_{fcc}$ + Ti-$O_b$ | Ti 3*d*+ Ti 3*p* + O 2*p* |
| E | 7,2 | Ti-$O_{fcc}$ | Ti 4*p* |
| F | 8.6 | Ti-$F_{fcc}$ | Ti 3*p* + F 2*p* |
| G | 11.3 | Ti-C | Ti 3*p* + C 2*s* |

[a] $O_{fcc}$ is O occupying the threefold hollow face-centered-cubic (fcc) site.
[b] $O_b$ is O occupying the bridge site between two surface Ti atoms.
[c] $F_{fcc}$ is F occupying the threefold hollow face-centered-cubic (fcc) site.

The intercalation of ions between 2D $Ti_3C_2T_x$ layers has been proposed as a promising approach for energy storage devices, such as batteries, capacitors, and hydride electrochemical devices, where different ions can be considered as a charge carrier [38]. However, the development of such devices requires a proper description of the $Ti_3C_2T_x$ surface, including correct identification of the termination species, surface site occupancies, and bonding configurations. The combined use of UPS, XPS, resonant PES, and valence band UPS/XPS computations presented in this work provides essential information that brings us closer to a complete understanding of MXenes and their unique characteristics important for future energy storage devices.

## 4. Conclusion

Chemical bond information between the $Ti_3C_2$-surface and the termination species in $Ti_3C_2T_x$ has been obtained using valence band UPS/XPS and resonant PES at low (27-80 eV) and high (780 and 1486.6 eV) photon energies. The study confirms that F and O sit on the threefold hollow face-centered-cubic (fcc) site. In addition, the study shows that a significant amount of O occupies a bridge site between two surface Ti atoms, both before and after F has been removed from the $Ti_3C_2$-surface. Furthermore, the comparison between experimentally obtained spectra and computed spectra shows that OH cannot be considered as an inherently formed termination species in $Ti_3C_2T_x$. The study reveals that the chemical bond between the $Ti_3C_2$-surface and the F on the fcc-site is dominated by the Ti $3p$ – F $2p$ hybridization. The chemical bond between the $Ti_3C_2$-surface and the O on the fcc-site is, on the other hand, dominated by the Ti $3p$ – O $2p$ hybridization with significant contributions of Ti $3d$ and Ti $4p$ states. The findings provide the necessary information to comprehend the unique characteristics of MXene and, thus, have implications regarding modeling of intercalation processes in, e.g., future energy storage devices.

## Acknowledgements

We acknowledge MAX IV Laboratory for time on Beamline SPECIES under Proposals 20190625, 20191107, and 20200582. Research conducted at MAX IV, a Swedish national user facility, is supported by the Swedish Research Council under contract 2018-07152, the Swedish Governmental Agency for Innovation Systems under contract 2018-04969, and Formas under contract 2019-02496. The computations were enabled by resources provided by the Swedish National Infrastructure for Computing (SNIC) at the National Supercomputer Centre (NSC) partially funded by the Swedish Research Council through Grant Agreement No. 2016-07213. We would also like to thank the Swedish Research Council (VR) LiLi-NFM Linnaeus Environment and Project Grant No. 621-2009-5258. The research leading to these results has received funding from the Swedish Government Strategic Research Area in Materials Science on Functional Materials at Linköping University (Faculty Grant SFO-Mat-LiU No. 2009-00971). M.M. acknowledges financial support from the Swedish Energy Research (Grant No. 43606-1) and the Carl Tryggers Foundation (CTS16:303, CTS14:310, CTS20:272). Most importantly, we thank Dr. Joseph Halim at Linköping University for preparing the samples.